\documentclass[12pt,preprint]{aastex}
\usepackage{natbib}
\shorttitle{Ellipticity Correlation with Atmospheric Structure}
\shortauthors{Asztalos et al.}

\begin{document}


\title{Properties of Ellipticity Correlation with Atmospheric Structure from Gemini South}


\author{S. Asztalos\altaffilmark{1}, W. H. de Vries\altaffilmark{2,3},
  L. J Rosenberg\altaffilmark{1}, T. Treadway\altaffilmark{1},
  \\D. Burke\altaffilmark{4}, C. Claver\altaffilmark{5},
  A. Saha\altaffilmark{5} and P. Puxley\altaffilmark{6}}

\email{asztalos1@llnl.gov}


\altaffiltext{1}{Lawrence Livermore National Laboratory, Livermore, CA 94551}
\altaffiltext{2}{University of California, Davis, CA 95616}
\altaffiltext{3}{Institute for Geophysics and Planetary Physics, 
Lawrence Livermore National Laboratory, Livermore, CA 94551 }
\altaffiltext{4}{Stanford Linear Accelerator Center, Menlo Park, CA 94025}
\altaffiltext{5}{National Optical Astronomy Observatory, Tucson, AZ}
\altaffiltext{6}{Gemini Observatory, Casilla 603, La Serena, Chile
Current address: National Science Foundation, Arlington, VA 22230}


\begin{abstract} Cosmic shear holds great promise for a precision
  independent measurement of $\Omega\rm_m$, the mass density of the
  universe relative to the critical density.  The signal is expected
  to be weak, so a thorough understanding of systematic effects is
  crucial.  An important systematic effect is the atmosphere: shear
  power introduced by the atmosphere is larger than the expected
  signal.  Algorithms exist to extract the cosmic shear from the
  atmospheric component, though a measure of their success applied to
  a range of seeing conditions is lacking.

  To gain insight  into  atmospheric shear,  Gemini South  imaging  in
  conjunction   with ground condition    and satellite wind data  were
  obtained.    We   find     that  under   good    seeing   conditions
  Point-Spread-Function  (PSF)  correlations persist   well beyond the
  separation typical  of high-latitude stars.  Under these conditions,
  ellipticity residuals based   on a simple  PSF interpolation  can be
  reduced to   within a factor   of a  few  of the  shot-noise induced
  ellipticity floor.  We also find  that the ellipticity residuals are
  highly correlated with wind  direction.  Finally, we correct stellar
  shapes   using a more   sophisticated  procedure and generate  shear
  statistics from stars.  Under all  seeing conditions in our data set
  the residual   correlations lie everywhere  below  the target signal
  level.   For  good    seeing  we find  that   the   systematic error
  attributable to atmospheric turbulence is comparable in magnitude to
  the statistical error (shape noise)  over angular scales relevant to
  present lensing surveys.
\end{abstract}





\section{Introduction}\label{intro}  Weak gravitational   lensing is a
powerful and  relatively unbiased technique  for measuring  the unseen
dark matter  and dark energy in the  universe \citep{BS01}. Light from
distant objects undergoes deflection  as it passes through intervening
over-dense  regions, inducing   tangential alignment  of  the objects'
image.  Numerous   cosmological parameters    can be    inferred  from
correlations in these alignments.  One  important lensing statistic is
the correlation in ellipticity between objects as  a function of their
separation.  For over  a decade this technique  has been used to infer
individual galaxy \citep{SDDS} and cluster  masses \citep{TYS}.   More
recently, the lensing  technique    has been extended   to   inferring
large-scale structure, where it is often referred to as cosmic shear
\citep{MW}.  Here  the   signal is weak   and  many  galaxies  must be
surveyed to overcome lack  of knowledge of the intrinsic ellipticities
(shape noise).    Nonetheless, this    technique has   been   yielding
estimates of $\sigma_8  \Omega\rm_m^{0.3}$ (the bias factor  times the
amount of  matter in the   universe relative to the critical  density)
consistent with other  techniques \citep{jarv06}.  The next generation
of  weak lensing instruments  will extend these  results to individual
redshift intervals.   From this, stringent  constraints will be placed
on  $w$ and  $w'$,  the cosmological equation of  state  and its first
derivative, respectively.

The   optics, camera  and  atmospheric  turbulence  all  contribute to
lensing systematic  errors, with the  latter effect being  perhaps the
most serious.   For a large   aperture telescope, a  thousand or  more
turbulent cells are in view at the aperture at  any given time, giving
rise to appreciable instantaneous   ellipticity.  For an  intermediate
length exposure of $\sim$ 15 s this raw ellipticity  is reduced by the
number of independent atmospheric realizations,  which in turn depends
on the wind speed.   Objects which happen to  lie close to one another
have  approximately sampled the   same portion of  the atmosphere  and
should have similar ellipticity  vectors; those further apart will  be
less similar.  In  this way the  atmosphere  induces correlations that
mimic a lensing signal.  Numerous  techniques  have been developed  to
remove spurious   atmospheric  power, however, these    techniques are
dependent on the spatial distribution of  fiducial objects whose point
spread functions (PSFs) are known.  Furthermore, these same techniques
may require a  degree  of coherence  between these objects,  otherwise
little  can be said  about the behavior of   the PSF between them.  In
short, atmospheric turbulence affects weak lensing measurements in two
ways: by introducing both   spurious ellipticity and causing   spatial
decoherence.  It is desirable   to characterize the magnitude  of this
effect in data and simulations.


The remainder of this paper is devoted to assessing the impact of
atmospheric turbulence on weak lensing measurements. Recently it was
shown that under good seeing conditions atmospheric residuals lie
comfortably below the level expected of a weak lensing signal
\citep{Witt05}.  The  need  for   controlling a  dominant
systematic  effect  like  atmospheric  turbulence, especially   in the
context of     future  ground based  lensing    measurements cannot be
understated.   Here  we systematically  study image  ellipticities and
build on that result by exploring image ellipticities  over a range of
seeing  conditions   with  known   atmospheric   turbulence  and  wind
conditions.  In Section
\ref{obs} we describe the image and  atmospheric data set.  In Section
\ref{anal}  we   show  that the  wind   direction  can be inferred  by
correlations  in  the    ellipticity  residuals.  We    then  turn  to
ellipticity  correlations as a function  of seeing and investigate how
seeing  affects PSF interpolation.    Photon shot noise introduces  an
irreducible ellipticity   floor   which is modeled   and  compared  to
observational data.  Finally, we construct lensing statistics from the
corrected ellipticities of PSF stars  and compare them with a putative
signal. These results may have bearing on  site selection for the next
generation of  proposed  telescopes and   also implications for  their
science requirements.

\section{Observations}\label{obs}

\begin{figure}[tb]
\epsscale{.95}
\plotone{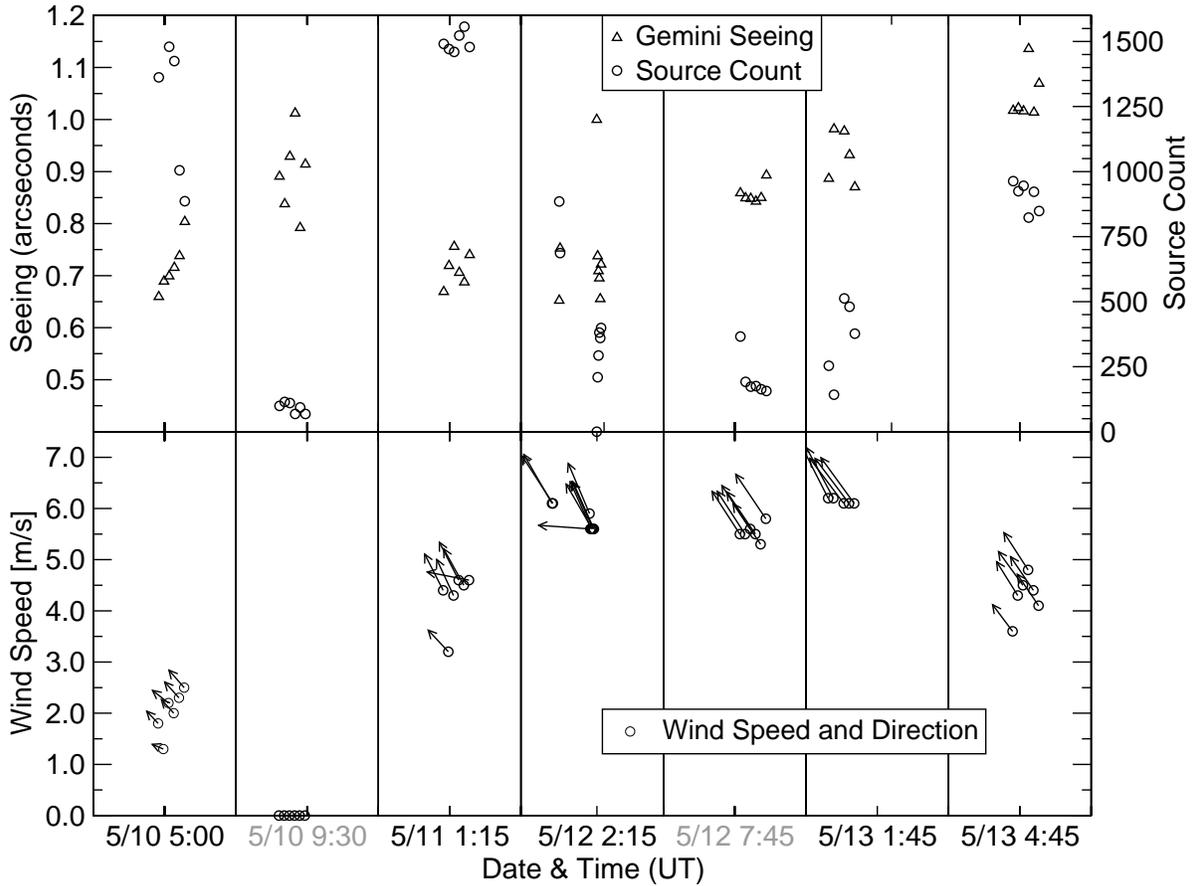} 
\caption{Forty-three images were acquired in two fields over four
  consecutive   nights.   The grey labels   along  the abscissa of the
  second and  fifth viewing windows   denote  the second field.  Image
  seeing  and source counts are  shown in the  upper panel; wind speed
  and direction in the lower panel.\label{DIMM}}
\end{figure}

To  study atmospheric impacts on   weak lensing science, discretionary
time was awarded on the Gemini Multi-Object Spectrograph (GMOS) Gemini
South \citep{Hook}.  GMOS is a three chip camera spanning 5.6$\arcmin$
on a side.  Forty three R-Band images were obtained over the course of
four  nights spanning May  10-13, 2005.   On  three of the four nights
(excepting 05/11) observations were divided into  two intervals, for a
total   of  seven viewing windows.     Observations  were done  on two
different fields  having bright  reference stars.  The  first (second)
field  is denoted with a  black  (grey) date  label in Fig \ref{DIMM}.
The  intrinsic stellar density  of the  first  field is  approximately
three times that of  the second. The  images were processed using  the
Gemini
IRAF\footnote{http://www.gemini.edu/sciops/data/dataSoftware.html}
reduction  routines  to  remove  bias and   gain  variation.  No  flat
fielding was  performed.  The upper plot shows  the median  seeing and
source object count, computed   with   the Gemini IRAF script     {\it
gemseeing}       and      SExtractor\footnote{http://terapix.iap.fr/},
respectively.  Good  seeing    prevailed during the   first  and third
viewing  windows; poor   seeing  prevailed during   the final  viewing
window.  Cloud  cover,  as inferred  by  the  source  counts,  was the
dominant effect in    the   remaining four  viewing    windows.  Rapid
excursions in the source counts throughout  are attributed to variable
cloud  cover.  The image seeing extracted  from images acquired during
the five remaining  viewing windows will  be used to  sort the results
into good,  median  and poor seeing bins.  

{\renewcommand{\baselinestretch}{0.90}
\begin{table*}[htbp]\begin{center}
\begin{tabular}{cccccccc} 
\hline
Frame&  WDIR    &   WS   &     DATE   &  UT   & AZ        &SO &    FWHM\\\hline
36  &   317.0   &   1.8  &     050510 &050106 &-147.91    &1436 &  0.718\\
37  &   293.0   &   1.3  &     050510 &050210 &-147.72    &1652 &  0.748\\
38  &   308.0   &   2.2  &     050510 &050313 &-147.53    &1543 &  0.753\\
39  &   318.0   &   2.0  &     050510 &050416 &-147.34    &1493 &  0.788\\
40  &   318.0   &   2.3  &     050510 &050519 &-147.16    &1112 &  0.818\\
41  &   319.0   &   2.5  &     050510 &050622 &-146.98    &1004 &  0.886\\\hline
43  &   333.0   &   4.4  &     050511 &011041 &146.25     &1553 &  0.726\\
44  &   317.0   &   3.2  &     050511 &011145 &146.42     &1535 &  0.788\\
45  &   335.0   &   4.3  &     050511 &011247 &146.60     &1525 &  0.825\\
46  &   333.0   &   4.6  &     050511 &011350 &146.77     &1582 &  0.766\\
47  &   331.0   &   4.5  &     050511 &011452 &146.95     &1613 &  0.748\\
48  &   281.0   &   4.6  &     050511 &011555 &147.13     &1542 &  0.809\\
52  &   330.0   &   6.1  &     050512 &011507 &147.70     &1003 &  0.710\\\hline
53  &   327.0   &   6.1  &     050512 &011610 &147.89     & 823 &  0.840\\
54  &   337.0   &   5.9  &     050512 &020037 &158.22     & 200 &  1.101\\
55  &   274.0   &   5.6  &     050512 &020140 &158.52     & 390 &  0.830\\
56  &   336.0   &   5.6  &     050512 &020242 &158.82     & 466 &  0.784\\
57  &   330.0   &   5.6  &     050512 &020345 &159.12     & 546 &  0.772\\
58  &   336.0   &   5.6  &     050512 &020448 &159.43     & 527 &  0.726\\
59  &   336.0   &   5.6  &     050512 &020550 &159.73     & 562 &  0.802\\\hline
74b &   333.0   &   6.2  &     050513 &013053 &151.69     & 430 &  0.993\\
75  &   332.0   &   6.2  &     050513 &013155 &151.92     & 329 &  1.088\\
77  &   323.0   &   6.1  &     050513 &013401 &152.41     & 665 &  1.080\\
78  &   323.0   &   6.1  &     050513 &013505 &152.65     & 636 &  1.029\\
79  &   324.0   &   6.1  &     050513 &013607 &152.90     & 542 &  0.965\\
87b &   323.0   &   3.6  &     050513 &044632 &211.57     &1074 &  1.136\\
88b &   328.0   &   4.3  &     050513 &044735 &211.77     &1039 &  1.165\\
89  &   324.0   &   4.5  &     050513 &044837 &211.96     &1058 &  1.140\\
90  &   327.0   &   4.8  &     050513 &044940 &212.16     & 947 &  1.324\\
91  &   326.0   &   4.4  &     050513 &045043 &212.35     &1037 &  1.145\\
92  &   326.0   &   4.1  &     050513 &045146 &212.54     & 970 &  1.242\\\hline
  \hline
\end{tabular}  \end{center}

\caption{Header data from 31 Gemini GMOS images on field 1 (13$^{\rm
    h}$40$^{\rm m}$40$^{\rm s}$ $-$53\arcdeg41\arcmin20\arcsec).   The
    first 6 columns are extracted directly from the image headers; the
    last two  columns are computed values.  The  headings from left to
    right refer to:   (Abbreviated)   frame  number, wind    direction
    (degrees), wind  speed (m/s),  date (yr\/mo\/dd), universal  time,
    telescope azimuth (degrees), source count  and full width at  half
    maximum (arcsec) defined as the median FWHM  of the source objects
    in the preceding column. All  angles are in degrees, North through
    East.\label{tb:DIMM}}
\end{table*}}

The lower plot in Figure
\ref{DIMM} depicts the  ground-wind speed and direction extracted from
the image   headers.  Ground-wind speeds   and  direction were roughly
constant, consistent   with  typical  site conditions.    The combined
effects of low intrinsic source count, limited number of exposures and
extensive cloud    cover make the  images  from  the  second and fifth
viewing  windows difficult to analyze,  hence  they are excluded  from
subsequent discussion.  A   summary  of header information  from   the
remaining frames is contained in Table
\ref{tb:DIMM}.  

\section{Analysis} \label{anal}

Galactic stars are unresolved and the photons they emit are relatively
unaffected  by   intervening    dark matter,  hence    are   excellent
zero-ellipticity     calibration objects.    However, the  atmosphere,
instrument and shot noise   all introduce spurious   ellipticity which
must be  removed  to recover  the lensing  signal.   The success  of a
procedure    in removing spurious  ellipticity  is  directly tested by
applying it to stars, whose  PSF one knows  a priori  to be round:  we
therefore limit our study to stellar objects.

Considerable variation in ellipticity definition is found in the
literature.  Here we adopt the definition

\begin{equation}
\epsilon = \frac{1-\beta^2}{1+\beta^2},
\end{equation}
\noindent in which $\beta$ is the ratio of the minor to major axes of
the fitted ellipse.  From this quantity one can construct the two
independent ellipticity components $\epsilon_1$ and $\epsilon_2$

\begin{eqnarray}
\epsilon_1  = & \epsilon\cos(2\phi)\label{er1}\\
\epsilon_2  = & \epsilon\sin(2\phi),\label{er2}
\end{eqnarray}

\noindent where $\phi$ is the angle  between major axis of the ellipse
and the   x-axis   of the detector.   Systematic     effects appear as
correlations  in ellipticity   vectors  across the image.   To  assess
quanitatively  the efficacy of  correction procedures for removing PSF
effects we define

\begin{eqnarray}
\epsilon_{RMS} &  = &
\frac{1}{N}\displaystyle\sum_{i}^N(\epsilon_{{1},i}^2 +
\epsilon_{{2},i}^2)^{1/2} \label{ser1}\\
\bar{\epsilon} &  = &
\frac{1}{N}\left[\left(\displaystyle\sum^N_i\epsilon_{{1},i}\right)^2 +
\left(\displaystyle\sum^N_i\epsilon_{{2},i}\right)^2\right]^{1/2}   \label{ser2}\nonumber \\ 
      &  = &(\bar{\epsilon}_1^2+ \bar{\epsilon}_2^2)^{1/2} \label{ser3}.  
\end{eqnarray}

\begin{figure}[tb]
\plotone{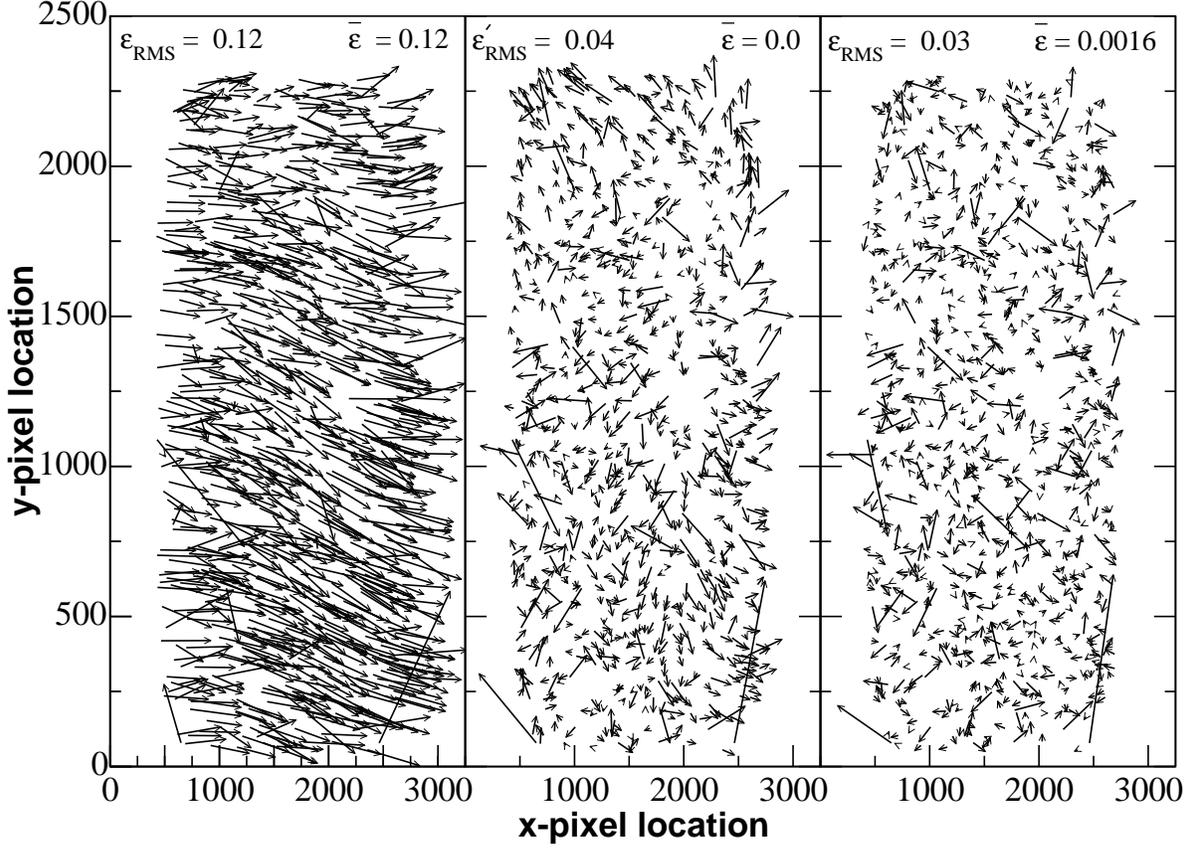} \caption{Raw ellipticities of 811 stars from frame 36
  (left) and after vector  subtraction of the average $\epsilon_1$ and
  $\epsilon_2$  ellipticities (middle).    From  its initial  value of
  0.12,  $\epsilon_{RMS}$ is reduced to  0.03 after processing through
  the lensing  pipeline (right).  More significantly, $\bar{\epsilon}$
  is reduced from 0.12 (left) to
  0.0016 (right).  Note that though (by construction) $\bar{\epsilon}$
  in  the middle plot is  zero, correlations nonetheless persist among
  the objects, especially near the edges.\label{S0036}}
\end{figure} 
  
\noindent The first quantity $\epsilon_{RMS}$ is the root-mean-square
ellipticity magnitude, while $\bar{\epsilon}$ is the average
ellipticity.  The sums in these equations are over all 811 stars.
Both quantities vanish in the round star limit; the latter likewise
vanishes if the ellipticity vectors are randomly oriented.

The  three panels in Figure \ref{S0036}  show the ellipticities of 811
stars in frame 36 derived using steps 1-3 of the measurement procedure
described in Section \ref{wla}.  (For reference, SExtractor identified
a total   of 1536 objects  in the  same frame.)   The  close agreement
between $\epsilon_{RMS}$  and $\bar{\epsilon}$  in the leftmost  panel
reflects the high degree of correlation (presumably telescope tracking
errors). A necessary next step  is the removal of spurious ellipticity
correlation, which otherwise mimics the weak  lensing signal.  A first
order correction is to remove the common mode seen in that panel.  The
middle panel in Figure \ref{S0036} is constructed in this manner, with
$\epsilon^{'}_{RMS}$ defined as

\begin{equation}
\epsilon^{'}_{RMS}=
\frac{1}{N}\displaystyle\sum_{i}^N((\epsilon_{{1},i} -
\bar{\epsilon}_1)^2+ (\epsilon_{{2},i} -
\bar{\epsilon}_2)^2)^{1/2} \label{ser4}
\end{equation}

replacing   $\epsilon_{RMS}$        of       Eqn.    \ref{ser1}    and
$\bar{\epsilon}_{1,2}$ defined as   in  Eqn. \ref{ser3}.  The   middle
panel  shows  the effect of  subtracting  off the lowest  order common
mode.   Though $\epsilon^{'}_{RMS}$ has  been  reduced to  0.04, there
remain persistent   correlations, especially near  the  image corners.
With  this  redefinition, $\bar{\epsilon}$ is   identically zero.  The
rightmost panel  in that  same figure  shows the  ellipticities  after
processing   the   image    through    the entire    pipeline.    Here
$\epsilon_{RMS}$ and $\bar{\epsilon}$ are  reduced to 0.03 and 0.0016,
respectively.  More important, no  obvious correlations  persist.  For
this  reason  we make extensive use   of the pipeline  when discussing
atmospheric residuals in Section \ref{wla}.

Having emphasized the preeminent role that stars play in subsequent
analyses and with some idea of the ellipticity magnitudes we now
consider how their correlations depend on wind and seeing conditions.

\subsection{Ellipticity Correlations with Wind}\label{wind}

The next generation of weak lensing measurements will likely include
high-fidelity simulations to disentangle systematic effects from the
signal.  A crucial component of these will be atmospheric simulations,
which require, among other things, wind speed and direction as a
function of height as their input.  These data may or may not be
readily available at many observing sites. Even if it does, it is
still non-trivial to infer an effective wind direction and speed
across the aperture. It therefore would be useful to extract wind
information from the image itself as one is accustomed to do for
seeing.

We see no wind information in the distribution of first moments of
object intensities (seeing).  However, it can be extracted by
considering the second moments (ellipticities).  The dominant
ellipticity trend observed in the leftmost panel in Fig.~\ref{S0036}
could well be attributed to the effects of wind transport across the
aperture, but may also contain non-atmospheric components such as
wind-shake of the telescope or guiding / tracking errors. We therefore
cannot assume a priori that the dominant ellipticity trend is solely
due to the wind.

\begin{figure}[tb]  \epsscale{1.0}   \plotone{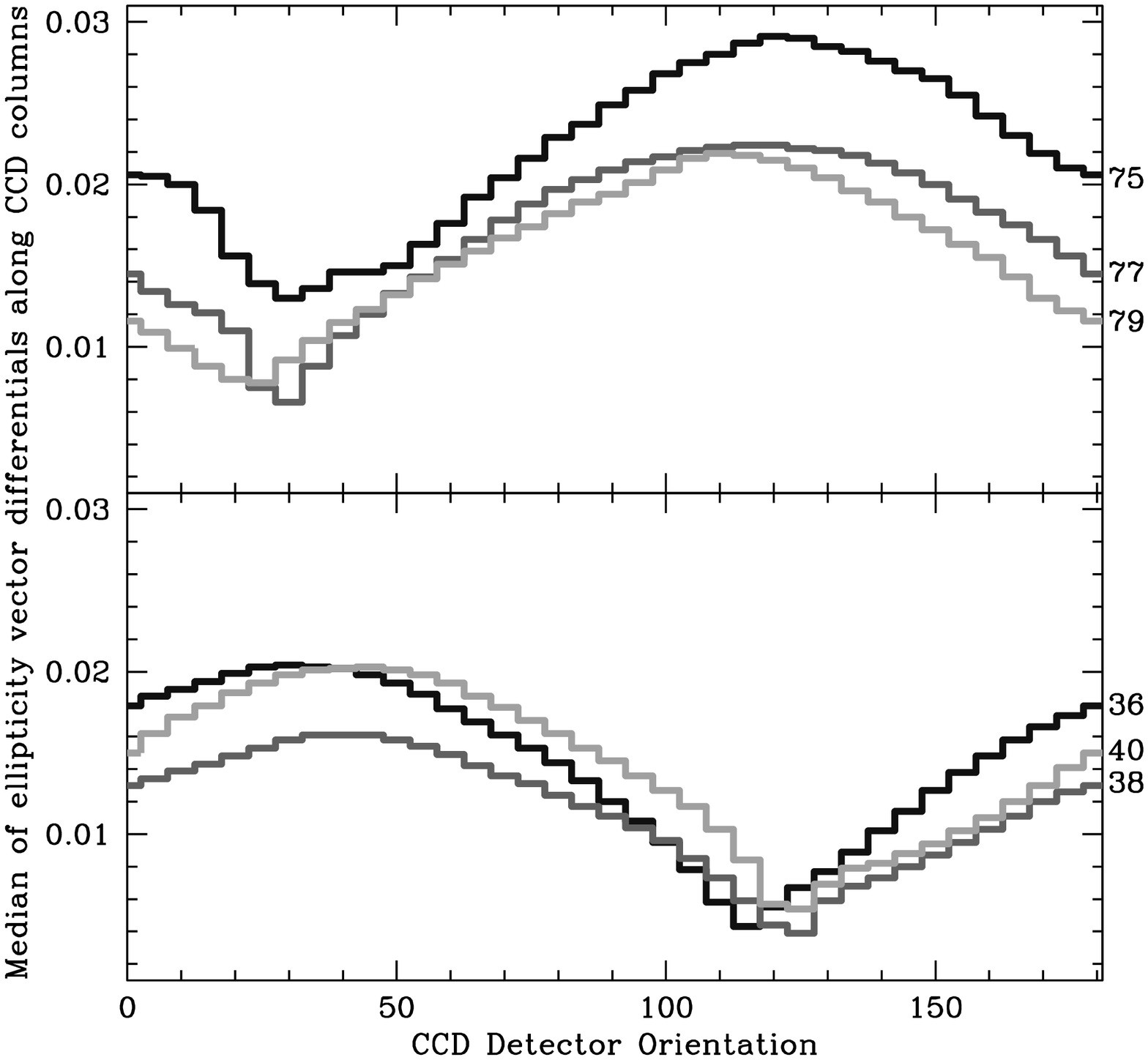}   
\caption{ Median value of the $\epsilon_y$ ellipticity residuals 
(see Eq. \ref{ser5})
of the pair-wise combination of N = 150 stellar objects per frame (for
a total of N(N-1) permutations), as a function of orientation angle of
the CCD detector. (The x-axis
  curve  (not   shown) would  appear   identical,  but shifted  by  90
  degrees.)  The frame   numbers are  indicated   on the  right.   The
  telescope was pointed approximately perpendicular to the ground wind
  direction in   the bottom set  of  three curves, and along  the wind
  direction  in  the top  three.  For each   of the frames a preferred
  position angle exists for which the projected residuals are smallest
  along the y-axis (and consequently largest  along the x-axis).  This
  position angle can  be  interpreted as  the direction for  which the
  wind induces the  longest correlation length, and therefore smallest
  residuals.  The median stellar separation is 163$\arcsec$, with 25\%
  of     the     pairs      having      separations     less      than
  105$\arcsec$.\label{windDir}}
\end{figure}

To   eliminate competing  phenomena as  the   origin  of the  dominant
ellipticity  we perform  pairwise    vector  subtraction of  the   raw
ellipticity components of stars across the image, with the ellipticity
components $e_a$ and $e_b$ derived from the shape parameter $\epsilon$
and position angle $\theta$ in the following manner
\begin{eqnarray}
\epsilon = &1 - {{e_a}\over{e_b}} \label{ellip}\\
\tan{\theta} = &{{e_b}\over{e_a}}. 
\end{eqnarray}
As  defined above, $e_a$ and  $e_b$ are the  lengths of the major- and
minor axes   of     the  ellipse, respectively.   It    follows   that
$\epsilon_{x}   =  \cos\theta$ and  $\epsilon_{y}=\sin\theta$ are  the
ellipticity  projections onto the  x-  and  y-axes  of  the  detector,
respectively.  The ellipticity residual is then defined as
\begin{equation}
\epsilon_{res} =  ((\epsilon_{{x},i} - \epsilon_{{x},j})^2 + 
(\epsilon_{{y},i} - \epsilon_{{y},j})^2)^{1/2}. \label{ser5}
\end{equation}
In constructing  this latter quantity  we arrive  at a distribution of
ellipticity residuals from which the  common-mode ellipticity has been
removed\footnote{We expect  these   non-atmospheric components  of the
ellipticity  to  be  extremely    uniform across  the   detector  and,
therefore,  to be    removed  very  efficiently.}.   The   ellipticity
residuals can be projected  onto the rows and  columns of the detector
(as  in Eq. \ref{ser5}),  along the RA and  DEC  coordinate system, or
along any  other set  of axes.   (For an  alt-az mount telescope  like
Gemini South a camera   rotator is used to  hold  the instrument at  a
fixed position angle.  For these data the position angle was such that
the axes  were aligned  with RA and  DEC.)   In Fig.~\ref{windDir}, we
plot the  magnitude  of  one  of  these  components as a   function of
position angle.  A position angle of  0\arcdeg\ corresponds to a basis
in which the x-axis corresponds  to the RA  direction, and the  y-axis
with    the DEC  direction.       Any   positive  rotation is     then
counter-clockwise through the East.   While the total magnitude of the
ellipticity is a rotational  invariant, the magnitude of the projected
values  are not.   Fig.~\ref{windDir} plots  the projection along  the
vertical axis;  by construction the  x-axis values are offset in phase
by 90 degrees.

The curve for each frame (as indicated on the right  side of the plot)
has a distinctive minimum.  Even though only a selected few frames are
shown,    all image data exhibit   this   behavior, albeit with  lower
significance for poorer seeing.  Because the ellipticity residuals are
at a minimum at a particular rotation,  we can interpret this position
angle   as  the  direction  in  which   the  residuals  are {\it most}
correlated: a longer  correlation length lowers the median ellipticity
value as the PSFs  are    more similar  along that  direction.    This
stretching  of the correlation   length  in a particular direction  we
believe  is due to the  more or less  constant wind  directions of the
dominant wind layers (ground-layer and 200 mb jet-stream layer) across
the telescope aperture.  In essence, this transport  of the air across
the telescope ensures  that  otherwise  uncorrelated regions  of   the
detector see the same atmosphere, resulting in more similar PSFs along
this effective wind direction.

During the observations the ground layer wind direction varied from
300 to 330 degrees (Fig. \ref{tb:DIMM}), with wind speeds of up to 6
m/s. The jet-stream layer was blowing due East with wind-speeds of
about 30 m/s. If we assume that the wind directions are constant, then
the only orientation variable left is the relative positioning of the
telescope with respect to these winds. 

\begin{figure}[tb]
\epsscale{1.0}
\plotone{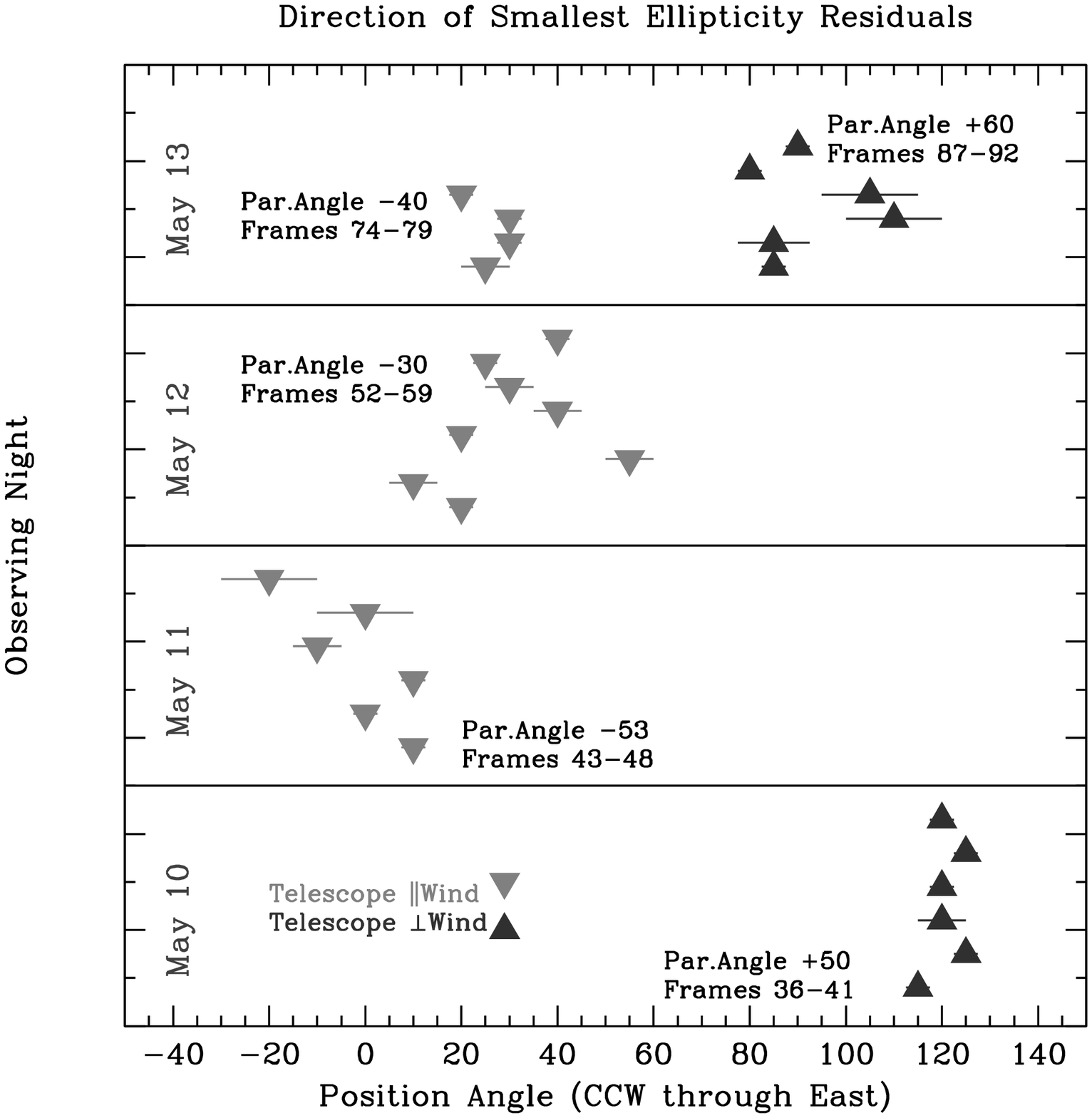}
\caption{Direction of smallest ellipticity residuals. The up (black)
  triangles   represent  frames  with   the telescope pointing roughly
  perpendicular to  the ground-wind; the  down  (grey) triangles where
  the   telescope is pointing along   the  ground-wind.  The telescope
  elevation is  about 60 degrees above the  horizon in all cases.  The
  position angle has been plotted from -40 to  140 degrees for display
  purposes. The horizontal   lines   are  indicative  of   measurement
  uncertainty.\label{windSummary}}
\end{figure}

Figure~\ref{windSummary}  summarizes the  minima  for   all our  image
frames.  It is clear  that the data fall  in two groups: one for which
the telescope  was pointed  along the  ground-wind  direction (azimuth
$\sim$150\arcdeg,   down     triangles),   and    one   pointed almost
perpendicular   to  the  ground-wind   (azimuth  $\sim215\arcdeg$,  up
triangles).   We  find that that    these minima are  consistent  over
multiple  days,  ultimately reflecting  similar  observation  and wind
directions.  The grey triangles  all reflect observations of  the same
field taken about 2 hours before the meridian, and the black triangles
2 hours after  the meridian.  However, given the  scatter in the data,
we cannot uniquely identify these minima with a particular wind layer.

We furthermore note that for each frame the minima, even after taking
out the bulk ellipticities through the pairwise subtractions,
correlate with the direction of the bulk ellipticity (as given by the
median ellipticity vector). The minima are found to be almost
perfectly at right angles with the direction of largest ellipticity,
and therefore closely align with the ellipticity minor axis. So, for
these particular Gemini observations it is clear that most (if not
all) of the bulk ellipticity observed in the frames is induced by the
wind.

Given the predominant wind direction at Cerro Pachon and other sites,
ellipticity residuals induced by the dominant wind layers are likely to
persist over time. If an observing strategy involved viewing a part of
the sky at similar hour-angles and telescope orientations, the result
will be a persistent residual that will not diminish by averaging more
images.

\subsection{Ellipticity Correlations with seeing}

In this section we describe how the atmosphere affects the ability to
characterize a PSF at a particular location on the detector, where
intrinsically round stellar objects can acquire an ellipticity for
reasons set forth in Section \ref{intro}. Obviously, the better we can
measure this unwanted ellipticity component the better we can correct
for it.

The ability to interpolate over the image depends, in part, on the
stellar density and atmospheric coherence.  Under perfect seeing (and
telescope) conditions a stellar wavefront is coherent and the PSF is
known everywhere.  Under finite seeing conditions the coherence
distance is also finite.  Coherence distances less than 1$\arcmin$ -
the approximate separation of high-altitude galactic stars - would
imply that there are locations in the image where the PSF is
substantially unknown.  This could impact weak lensing analyses, which
relies on stellar objects to predict the PSF of galaxies.  For this
reason one would like to first establish the coherence distance as a
function of seeing.

\begin{figure}[tb]
\plotone{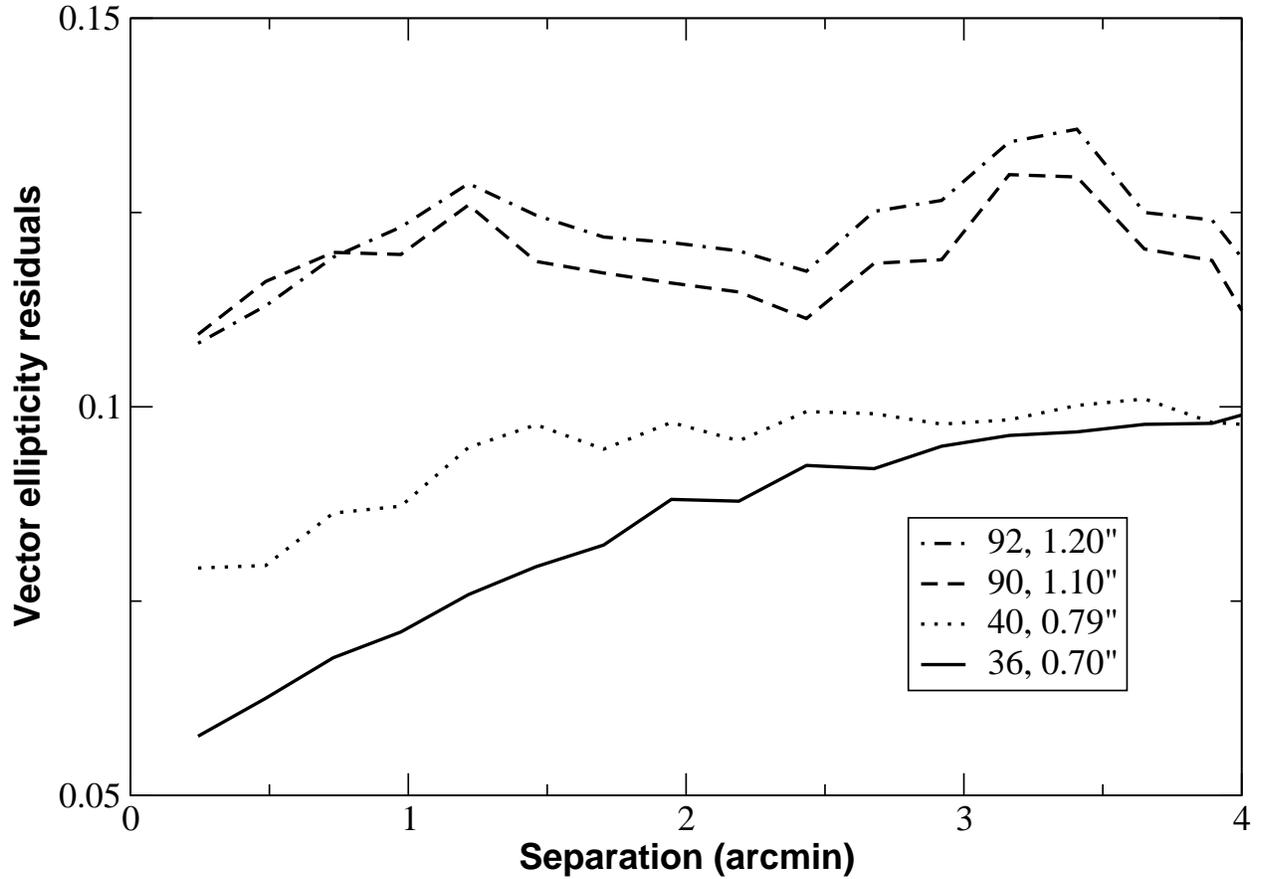}
\caption{(left) Vector ellipticity residuals constructed from the 400
  brightest   stars of   four   images  spanning  a range  of   seeing
  conditions.   The  residuals were placed   into 27 bins separated by
  $\sim$ 0.25$\arcmin$   out  to  a  maximum  separation    of  $\sim$
  6.5$\arcmin$,  though statistics  limit  the  useful range to  below
  4$\arcmin$.     From    each    bin   the    median   residual   was
  extracted.\label{cohere}}
\end{figure} 

Figure \ref{cohere} shows the vector  ellipticity residuals of the 400
brightest stars as  a function of separation  for four frames spanning
the range of seeing  conditions in Table \ref{tb:DIMM}.  Residuals are
again constructed\footnote{With e$_1$ and e$_2$ as defined in Eqs
\ref{er1} and \ref{er2} now replacing e$_a$ and e$_b$ in Eq \ref{ser5}.} 
for each frame and placed into bins separated by $\sim$ 0.25$\arcmin$.
The median  residual was extracted  and plotted.   For good seeing the
residuals are small for  small separations.  These residuals grow with
increasing separation,  reflecting  atmospheric  decoherence on  these
scales.  This trend becomes less pronounced as  the seeing gets worse.
By the time  the seeing  is $\geq 1\arcsec$   the atmospheric is  largely
decoherent at all separations.

This  analysis addresses the PSF  behavior   at stellar locations  and
demonstrates that under good seeing conditions the atmosphere exhibits
some  degree of  coherence for  small   separations.  To  explore more
quantitatively the useful limit of seeing we turn  to the issue of the
PSF behavior at an arbitrary location within the image.  We choose two
sets of stellar objects: fiducial and fixed.  In what follows the PSFs
of  fiducial objects are  measured and corrections  are applied at the
location of the fixed set.  The density of  the fiducial set is varied
by selecting progressively fewer stars.  It  should be noted that this
selection is not completely  random:  we always include  the brightest
objects  (i.e., the density is  lowered by removing the faintest stars
first).  This ensures that  we are using the  brightest and easiest to
measure  stars,  mimicking the   lensing  pipeline  methodology.  This
method also minimizes  the impact of photon  shot noise limitations to
the ellipticity (see  \S~\ref{sh}).    Using these fiducial  stars  we
predict  the PSFs  of   the other fixed  set  of  stellar  objects  by
calculating the (distance) weighted mean PSF derived from the fiducial
set.

\begin{figure}[tb]
\epsscale{1.0} 
\plotone{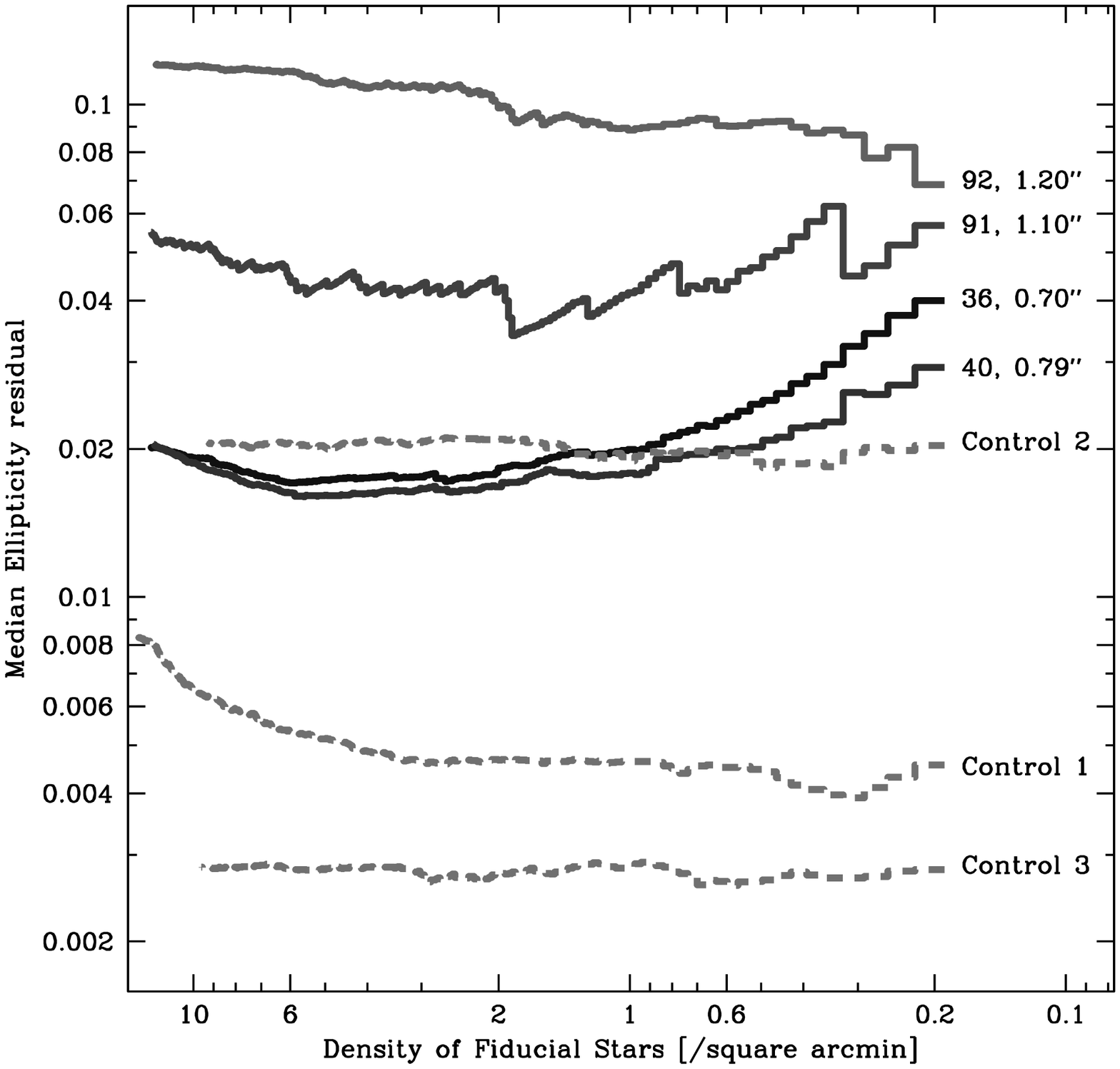} 
\caption{PSF interpolation as function of decreasing fiducial star spatial 
  densityfor four selected images of the same field.  The frame number and
  corresponding median seeing are shown.  The control curves (marked 1
  through 3) are based on artificial images containing round stars
  (0.7 $\arcsec$ seeing) selected from a brightness distribution of
  frame 36 (1), 20$^{\rm th}$ magnitude stars only (2), and 15$^{\rm
    th}$ magnitude stars only (3).  Instances of good seeing
  ($<0.8\arcsec$) allow for excellent interpolation of the PSF down to
  densities of a few per square arc minute.  The interpolation is
  limited by photon shot noise induced ellipticity, as illustrated by
  the dashed control~2 histogram.  The upturn toward high spatial
  densities as seen in frames 36, 40, and control~1 is due to the
  progressive inclusion of fainter PSF objects in order to attain the
  required spatial density.  \label{psfmap}}
\end{figure}

We use a  fixed set of 150 stellar  PSF positions randomly distributed
across the   chip.    For each   of  these  we  calculate  the  vector
ellipticity  difference  between the    distance  weighted mean   PSF,
calculated using  a varying set of fiducial  stars, and the actual PSF
at these 150 positions.  The y-axis of Fig.~\ref{psfmap} is the median
value ellipticity residual for the fixed set of 150 stars.  The number
of  fiducial stars comprising  the interpolation grid  are varied from
about 400 down to about 5, corresponding to a spatial density range of
$\sim14$  down to 0.2 stars per  square arc minute (given our detector
surface area of 29.2 square arcminutes).

In Fig.~\ref{psfmap}  we plot the results for  frames 36,  40, 91, and
92,  whose variables  span   the range  of  observing conditions  (cf.
Table~\ref{tb:DIMM}).  The curve based on frame 36 represents the best
observing conditions ($\sim0.7\arcsec$  seeing), and will be discussed
first.  This curve is flat across spatial densities between 10 down to
2 stars /  $\square\arcmin$, only to turn upward  at higher and  lower
spatial densities.  The upturn  with decreasing spatial density is due
to the gradual onset of spatial decorrelation  between the PSFs due to
the atmosphere.  With separations  of more than a  few arc minutes the
simple  weighted interpolation  invoked   here poorly  estimates   PSF
shapes. The upturn in the  left part of the curve,  on the other hand,
is due   to probable contamination  by non-stellar  objects (and their
large intrinsic  ellipticities) of the  interpolation grid, as well as
ellipticity  measurement   limitations of   faint  stars due  to  shot
noise. We come back to this issue shortly.

Ignoring this high  spatial density upturn, it is  clear that with  an
interpolation grid of at least 1 star / $\square\arcmin$ residuals are
minimized    within  this type of     interpolation   scheme.  As  the
observation  conditions worsen (as they do  through frames 40, 91, and
92) ellipticity residuals   rise  regardless of PSF surface   density,
reproducing the trend seen  in Fig.~\ref{cohere}.  For the (seemingly)
pathological cases of frames 91 and  92, for which the PSF correlation
length is essentially zero, we do not see an increase in the residuals
with decreasing PSF surface density.  Instead, we see a {\it decrease}
due  to the fact  that in  the  lowest spatial  density  bins we  only
consider the brightest stars in the frame, which happen to be the ones
with  the smallest measurement  errors.  The spatial PSF decorrelation
in these  frames is such  that it does not matter  what PSF we use for
our model.

\subsubsection{Shot Noise contributions}\label{sh}

To address how {\it well} this interpolation scheme performs,
artificial sky images were created using a realistic observing and
instrumentation setup. We use SkyMaker
software\footnote{http://terapix.iap.fr/} to create three sets of
images.  The first (labeled ``control 1'' in Fig.~\ref{psfmap}) uses
the brightness distribution found in frame 36. The other two images
contained only stars of 20$^{\rm th}$ and 15$^{\rm th}$ magnitude
(control 2 and 3 respectively). All stars are generated with a FWHM of
0.7 $\arcsec$ but no optical distortions are applied, hence the PSFs
are perfectly round.  Consequently, the sole source of ellipticity is
photon number statistics, i.e., shot noise.  Furthermore, as we have
perfect PSF correlation across the chip, there should not be any trend
with decreasing spatial PSF density.  This is indeed what is seen in
both controls 2 and 3 and the offset between these histograms is
exactly what is expected from shot noise calculations.  The fact that
control~1 slopes upwards is due to the inclusion of progressively
fainter stars as one goes up in spatial density (the faintest stars in
control~1 are 19$^{\rm th}$ magnitude), which increases the shot-noise
contribution.

Based on these  control curves  we can  conclude  the following.  With
ellipticity residuals on  order of  0.02 (frame 36),  we are  within a
factor of  a few of  the  shot noise limit   for $\sim$ 19$^{\rm  th}$
magnitude   stars, with   the   difference  attributed to  atmospheric
turbulence.  Second, for stellar surface  densities of less than about
$1 /
\square\arcmin$ this interpolation scheme is less successful. However,
the roll-off is gradual, both in terms of PSF surface density and
seeing conditions.

\subsection{Full lensing analysis}\label{wla}

Ellipticity correlations as formulated above are a powerful measure of
coherence and PSF  behavior, but  are not  expressed  a form  that  is
readily compared with   a cosmological lensing signal.  Further,  more
precise correction  procedures are available  than that applied above.
We have chosen the Bernstein and Jarvis weak lensing analysis software
\citep{BJ} to correct 
stellar shapes  for systematic effects.   Much of the shape correction
procedure is similar to   that described in  \citep{Jarv03}.  However,
not  all elements of a pipeline  expressly designed  to measure galaxy
shapes apply here, where we restrict our attention to stellar objects.
Those that do apply are described briefly below.

\begin{enumerate}
    \item{{\it Object Identification.} All our analyses adopt SExtractor
    as their starting points.  Aperture magnitudes and the zero-point
    magnitude suggested for GMOS are adequate for our purposes.  Only
    well isolated stars having no associated error flags are selected.}

  \item{{\it Shape measurement.} Ellipticities measured by SExtractor
    serve as initial estimates for more precise shape measurements
    based on second moments weighted by elliptical Gaussians.}

  \item{{\it Star Identification.} Stars are discriminated from
    galaxies   on  the basis  of   size-magnitude  considerations.  To
    minimize    the   shot-noise   contribution    and  further reduce
    contamination we selected the $N$ brightest candidates.}

  \item{{\it Convolution.}  A fiducial subset of these stars is chosen
  for  rounding.   The kernel required  to round  this  subset is then
  interpolated to all other image locations. No correction is made for
  PSF dilution as described in \citep{Jarv03}.}

  \item{{\it Shape Remeasurement.} With all stellar objects either
    directly or indirectly rounded via interpolation the shapes are
    remeasured. We do not correct for centroid bias, nor do we combine
    shape measurements.}
\end{enumerate} 

\begin{figure}[tb]
\plotone{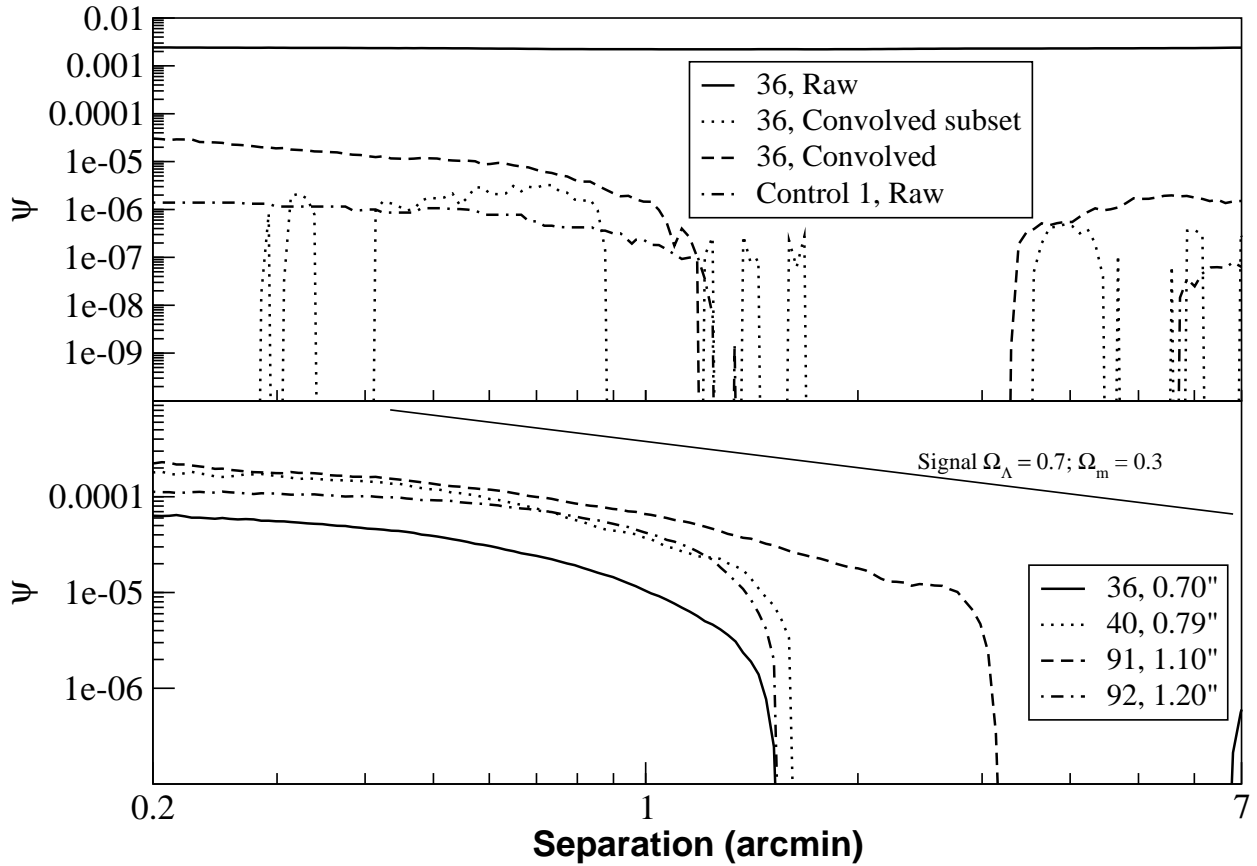} 
\caption{[Upper panel] The two point correlation function $\psi$ for
  various object lists processed through   the pipeline.  The   solid,
  constant, line  is  the  correlation function   for 512 raw  stellar
  objects  from frame   36.  That the  curve   is featureless and  the
  amplitude large   is   a  reflection  of  the    sizable ellipticity
  correlation displayed in the   leftmost panel in Fig.   \ref{S0036}.
  The dotted curve is constructed from a subset of 122 rounded stellar
  objects ($\sim$  25\% of the  total of  512).   The erratic behavior
  observed in  this curve arises  from  minimal statistics in numerous
  separation bins.  The dashed curve is again the correlation function
  for the 512  stars, but now the  122 stellar objects have  been used
  for  rounding and  interpolation.    The dashed-dotted curve is  the
  correlation function of    339  simulated stars that   comprise  the
  ``Control 1''  group   from Fig.  \ref{psfmap}.  [Lower  panel]  The
  two-point correlation  functions  for frames   36,  40,  91  and  92
  constructed in manner similar to the dashed curve in the upper plot,
  but instead using 30 of the  brightest stars as the fiducial subset.
  Only positive values  of $\psi$ are   shown.  The solid line  is the
  level of the expected signal for a cosmology with $\Omega_{\Lambda}$
  = 0.7, $\Omega_m$ = 0.3 and z = 1 \citep{JS}.\label{pipe}}
\end{figure}

A similar but  distinct analysis has  been carried out with data  from
the    Subaru    telescope   on a     single   atmospheric realization
\citep{Witt05}.  There it was  found that atmosphere-induced residuals
are a factor of  $\sim$ 20 or so  below  the the target signal  level.
Here we extend  that study to a range  of atmospheric  conditions. The
upper panel in Figure
\ref{pipe} shows the two-point correlation $\psi$, defined as
\begin{equation}
\psi = (\epsilon_1\epsilon_1 + \epsilon_2\epsilon_2)/4
\label{xi}
\end{equation}

for  various object lists  processed through the  pipeline.  The solid
line is  the  correlation function  constructed  from  512 uncorrected
stars from Frame 36,  resulting in a constant  $\psi \sim 0.002$ line.
The lack  of features in this curve  reflects the dominant ellipticity
trend evident in the leftmost panel in  Fig.  \ref{S0036}.  The dotted
curve is constructed from a fiducial subset of  122 rounded stars from
that same  frame.  The dashed curve is  again the correlation function
of the same 512 stars,  but with the fiducial  subset of 122 stars now
used for rounding  and interpolation.  The  dashed-dotted curve is the
correlation   function constructed  from    339 simulated stars   that
comprise the  ``Control 1'' group  of Fig.   \ref{psfmap}. Recall that
``Control 1'' can  be  considered the same as   frame  36 but  without
atmospheric distortions.  The  correlation functions generated in this
manner are bounded from  above by the  raw ellipticities (black curve)
and from below by photon shot noise (dash-dotted curve), as expected.

The  lower panel in  Figure  \ref{pipe} was generated  in an analogous
manner to that used to produce the dashed curve in the upper plot, but
is  carried out for  frames  36, 40, 91  and  92 using just  30 of the
brightest stars in each frame as fiducial  stars.  The convolution was
then applied  to the remaining stars.   The roll off exhibited by each
of the curves for separations above a few  arcseconds is understood in
terms of the density  of fiducial stars ($\sim$  1/$\square\arcmin$ to
match the typical density   of high-latitude stars).   For separations
below a few arcseconds the correlations rise monotonically, reflecting
the imperfect interpolation process  between fiducial stars.  However,
even for the    smallest  separations this procedure  results    in an
improvement on  systematics   shape 3-10  times over that   in the raw
images.  Under  all  seeing  conditions and   for  all separations the
uncorrectable power $\psi$  is at worst a factor  of five  or so below
that of  the  expected lensing   signal  \citep{JS}.  For  separations
relevant to present surveys ($\sim$ 1$\arcmin$ or more) the systematic
error arising  from uncorrectable atmospheric  power (for good seeing)
is  of  comparable magnitude  to  the statistical error ($\sim$ 0.003)
\citep[Fig.1]{jarv06}.   The  lack  of  an   exact one-to-one  correspondence
between seeing and amplitude of the two-point correlation functions is
attributed  to fluctuations, as  well as  variations in  the number of
test stars in the four frames.

\section{Conclusions}

Many systematic effects will  affect data from  the next generation of
weak lensing  instruments.   To understand the   impact of the varying
atmosphere we have  acquired simultaneous  image and atmospheric  data
from  the Gemini South  telescope.  The  exposure  time of 15  s, site
location and mirror diameter are matched to that of the proposed Large
Synoptic Survey Telescope \citep{LSST}.

We  have studied  the effects  of  atmospheric turbulence  using three
methods of increasing  sophistication and demonstrate that with seeing
conditions   up to  $\sim$  0.7$\arcsec$  the  PSF   is coherent  over
appreciable  angular scales.   When   seeing exceeds   1$\arcsec$, the
atmosphere  is largely decoherent over  all  angular scales.  We  find
that  these  correlations  are  not isotropic.    Rather, there   is a
preferred direction,  which we find to be  induced by  persistent wind
layers.  Unless somehow  corrected for, this  correlation will persist
even after image stacking.  Using a simple interpolation technique and
under  good seeing  conditions the PSF  can be  predicted to within  a
factor of  a  few of the  theoretical (shot  noise) limit.  A complete
weak lensing analysis demonstrates that for data taken under less than
ideal circumstances atmospheric residuals can be reduced below that of
the target cosmic shear  signal.  The presence of  fiducial stars at a
nominal density of  $\sim$ 1/$\square\arcmin$ provides for  sufficient
sampling of the image plane to compensate for atmospheric decoherence.
Under the best seeing conditions we confirm an earlier result
\citep{Witt05} and further note that the uncorrectable systematic errors 
are of the order of statistical errors associated with present lensing
surveys.   However, future surveys will  measure  many more shapes and
the statistical  errors  will be correspondingly  reduced.   Thus, the
present level of systematic error  must also be lowered, especially in
the  non-linear regime.  A  priori, it  is not  obvious  how one would
commensurately lower the systematic error attributed to the atmosphere
using  the procedure  outlined in  this  paper  (since the errors  are
proportional to the  fixed  density of stellar   objects used to  make
shape corrections).  Fortunately, future surveys will revisit the same
patch  of    sky multiple  times and  one   can resort  to shear-shear
correlations computed from shear measured across frames
\citep[first discussed in][]{Jarv04}.   If these frames are  well 
separated temporally (e.g. $>$  20 s) then  the atmospheric noise is
purely  stochastic  and  will   not  add  systematic   power to  shape
measurements
\citep[see][for a methodology that minimizes the information lost
from not computing shear correlations within an image]{jjb06}.

\acknowledgments{ This research is supported by the U.S. Department of
Energy  under  contracts W-7405-ENG-48.  We  would like  to thank Mike
Jarvis for   his  assistance  with  the  analysis  pipeline  and error
estimation.  We acknowledge fruitful discussions with Steve Kahn, John
Peterson, Garrett Jernigan, Scot Olivier and Dave Wittman.  We further
wish to  thank Steve Heathcote  for his hospitality  while at the CTIO
site.   The comments from the  referee were  most helpful in improving
the paper.

Based on observations obtained at the Gemini Observatory, which is
operated by the Association of Universities for Research in Astronomy,
Inc., under a cooperative agreement with the NSF on behalf of the
Gemini partnership: the National Science Foundation (United States),
the Particle Physics and Astronomy Research Council (United Kingdom),
the National Research Council (Canada), CONICYT (Chile), the
Australian Research Council (Australia), CNPq (Brazil) and CONICET
(Argentina), Program GS-2005A-DD-9.
}



{\it Facilities:} \facility{Gemini:South (GMOS)}



\bibliography{ms}

\begin{thebibliography}{13}
\expandafter\ifx\csname natexlab\endcsname\relax\def\natexlab#1{#1}\fi

\bibitem[{{Bartelmann} \& {Schneider}(2001)}]{BS01}
{Bartelmann}, M. \& {Schneider}, P. 2001, {Physics Reports}, 340, 291

\bibitem[{{Bernstein} \& {Jarvis}(2002)}]{BJ}
{Bernstein}, G. \& {Jarvis}, M. 2002, AJ, 123, 583

\bibitem[{{Fischer} {et~al.}(2000)}]{SDDS}
{Fischer}, D. {et~al.} 2000, AJ, 120, 1198

\bibitem[{{Hook} {et~al.}(2006)}]{Hook}
{Hook}, I. {et~al.} 2006, PASP, 116, 425

\bibitem[{{Jain} \& {Seljak}(1997)}]{JS}
{Jain}, B. \& {Seljak}, U. 1997, AJ, 560

\bibitem[{{Jain} {et~al.}(2006)}]{jjb06}
{Jain}, B. {et~al.} 2006, Journal of Cosmology and Astroparticle Physics, 1

\bibitem[{{Jarvis} \& {Jain}(2004)}]{Jarv04}
{Jarvis}, M. \& {Jain}, B. 2004, astro-ph/0412234

\bibitem[{{Jarvis} {et~al.}(2003)}]{Jarv03}
{Jarvis}, M. {et~al.} 2003, AJ, 125, 1014

\bibitem[{{Jarvis} {et~al.}(2006)}]{jarv06}
---. 2006, ApJ, 644, 71

\bibitem[{{Mellier} \& {van Waerbeke}(2001)}]{MW}
{Mellier}, Y. \& {van Waerbeke}, L. 2001, in Where's the Matter?, ed.
  M.~{Treyer} \& L.~{Tresse} (Frontier Group)

\bibitem[{{Tyson} {et~al.}(1990)}]{TYS}
{Tyson}, J. {et~al.} 1990, ApJL, 349, 1

\bibitem[{{Tyson} {et~al.}(2005)}]{LSST}
---. 2005, Bulletin of the AAS, 207

\bibitem[{{Wittman}(2005)}]{Witt05}
{Wittman}, D. 2005, ApJL, 632, 5

\end{thebibliography}

\clearpage



\clearpage




\end{document}